\documentclass[11pt]{article}

\pdfoutput=1

\usepackage[sfmath]{kpfonts}

\usepackage[T1]{fontenc}
\usepackage[utf8]{inputenc}
\usepackage{xspace}

\usepackage{url}
\usepackage{hyperref}
\usepackage{graphicx}
\usepackage{subcaption}
\usepackage{xcolor}
\usepackage{comment}

\title{Exploring term expansion for task-based retrieval at the TREC-COVID track}

\author{
    Thomas Schoegje\footnote{Utrecht University, Utrecht, the Netherlands} \\
    \texttt{\small t.schoegje@utrecht.nl}
    \and 
    Chris Kamphuis\footnotemark[2] \\
    \texttt{\small c.kamphuis@cs.ru.nl}
    \and
    Koen Dercksen\footnote{Radboud University, Nijmegen, the Netherlands} \\
    \texttt{\small koen@cs.ru.nl}
    \and
    Djoerd Hiemstra\footnotemark[2] \\
    \texttt{\small hiemstra@cs.ru.nl}
    \and
    Toine Pieters\footnotemark[1] \\
    \texttt{\small t.pieters@uu.nl}
    \and
    Arjen P. de Vries\footnotemark[2] \\
    \texttt{\small arjen@acm.org}
}
\date{October 2020}

\newcommand{\todo}[1]{\textcolor{red} #1}

\begin{document}
\maketitle

\begin{abstract}

We explore how to generate effective queries based on search tasks. Our approach has three main steps: 1) identify search tasks based on research goals, 2) manually classify search queries according to those tasks, and 3) compare three methods to improve search rankings based on the task context. The most promising approach is based on expanding the user's query terms using task terms, which slightly improved the NDCG@20 scores over a BM25 baseline. Further improvements might be gained if we can identify more specific search tasks.
\end{abstract}

\section{Introduction}
The COVID-19 pandemic created new information needs around the virus. The TREC-COVID track \cite{DBLP:journals/corr/abs-2005-04474} is a response by the Information Retrieval community, that investigates how we can better serve these needs. Generating effective queries was found to be an important strategy during early rounds of this track \cite{DBLP:journals/corr/abs-2007-07846}. The input for this query generator uses terms that may not be available during a real life ('naturalistic') search process. We explored an alternative query formulation based on search tasks.

In our approach, we group the search topics into typical search tasks, which in turn provide context for improving search results. The three main steps in our approach are 1) to identify the search tasks, 2) to classify search queries into search tasks, and 3) to improve search results based on the task context.

The first step is to identify potential search tasks, which we base on the key COVID-related research goals that were identified by a number of organizations.


The second step is to classify search queries into tasks. We compare automatic classification to manual annotation, and, due to low automatic prediction accuracy, settle on manual annotation for our approach.

Finally we change search rankings based on the task context. Three methods are introduced and compared to bm25 (Anserini) baselines. The most promising method is a query generation approach based on task term expansion, although it did not yet yield significant improvements over the baseline. The second re-ranks search results based on doc2vec, and the third re-ranks based on the publication journal.


Using this approach, we explored task-based query expansion with three research questions:

\begin{description}
    \item[RQ1] Does term expansion with task terms improve results compared to using only query terms?
    \item[RQ2] Does term expansion with task terms further improve the query generated by the University of Delaware's query generator?
    \item[RQ3] Is the task categorization able to represent the topics introduced in later rounds?
\end{description}

The expanded query in RQ2 refers to the successful approach to query formulation found during early rounds on the TREC-COVID track \cite{DBLP:journals/corr/abs-2007-07846}. These are taken from the TREC search topics' question attribute, which is a sentence that describes the search topic's information need in natural language. RQ2 tests if the tasks and the topic query terms are independent sources of information.

In the following Section we give background information on the TREC-COVID track. Section 3 contains related work to our approach, which is introduced in Section 4. During Section 5 we present our experiments and results. We present our conclusions in Section 6. As our approach was rather explorative, we report with some of the negative results and dead ends in Section 7.


\section{Background}

\subsection{CORD-19 dataset}
The CORD-19 dataset is a collection of scientific literature on the coronavirus \cite{DBLP:journals/corr/abs-2004-10706}\footnote{Available at \url{https://www.semanticscholar.org/cord19}}. It was created by the Allen Institute for Artificial Intelligence to stimulate data science research related to the virus  The CORD-19 dataset is a collection of scientific literature surrounding the coronavirus. At the time of publication daily versions of this dataset are being generated by performing a COVID-related query within a number of repositories of medical literature. These include preprint repositories (e.g. BioRXiv) and peer-reviewed repositories (e.g. PubMed). Metadata and full-text access were gathered for most documents. Some clustering was performed to de-duplicate results. 


\subsection{TREC COVID track}
The TREC-COVID challenge organized the IR research by creating search topics and evaluating search rankings. This was done over the course of five rounds, each of which introduced new topics and provided evaluations of the newest search rankings. A diverse set of simulated search topics were developed based on the diverse and evolving information needs of users during the pandemic \cite{roberts2020trec, DBLP:journals/corr/abs-2005-04474, disco}.

Each search topic includes three fields: a \emph{query}, the \emph{question} a user is trying to answer (a natural language sentence), and a slightly longer \emph{narrative} giving context on the search topic. 

Search rankings, known as runs, were evaluated each round by taking the top-ranked results in a run and asking people with some domain expertise whether a given document was indeed relevant to the search topic. These relevance assessments were collected for all runs. Each research team was allowed three runs in each round and five additional runs in the final round. In order to avoid the situation where teams prioritize documents that are known to be relevant, the evaluation each for round used only the residual dataset (the dataset without the previously judged documents).

\section{Related work}
One of the most influential papers in identifying the goals behind search was Broder's taxonomy of web search \cite{DBLP:journals/sigir/Broder02}. Rose and Levinson identified three steps necessary for supporting search results based on such tasks \cite{DBLP:conf/www/RoseL04}. The first step is to identify the search tasks. Previous work has typically identified tasks based on some form of query log analysis (e.g.  \cite{DBLP:journals/sigir/Broder02, DBLP:conf/www/RoseL04}). The second step is to associate these tasks with the search queries that were issued.  V{ö}lske et al. compared a number of methods for mapping queries to tasks, and found that an inverted index-based method performed best, with an accuracy over 60\% \cite{DBLP:conf/sigir/VolskeFSH19}. In this approach the tasks are indexed in a small search engine, and the classification occurs by retrieving the top result for a given query. Over the course of a search session, it may also be possible to use search behavior signals to identify user tasks \cite{DBLP:conf/sigir/MitsuiS19}.

The final step to task-based ranking according to Rose and Levinson is to exploit the search task information to improve results. Early rounds of the TREC-COVID track proved term expansion to be a valuable tool for improving search results \cite{DBLP:journals/corr/abs-2007-07846}. The new terms were based on the context that was added to each query in the TREC topics, specifically the topic's question attribute. This approach extracted all biomedical named entities in the topic query and topic question using ScispaCy \cite{DBLP:conf/bionlp/NeumannKBA19}, and used these terms as the new query.






\section{Method}
Improving search results based on knowledge of the user's search tasks involves three main steps:
\begin{enumerate}
    \item Identifying the potential search tasks
    \item Mapping the user queries to these tasks
    \item Improving search rankings based on their search tasks
\end{enumerate}

\subsection{Identifying tasks}
Potential search tasks in this domain need to be identified and gathered in a task framework. We base these tasks on the research goals that have been identified to deal with the pandemic. One source of research goals are the tasks in Kaggle's "COVID-19 Open Research Dataset Challenge" , which was published alongside the original CORD-19 dataset \cite{DBLP:journals/corr/abs-2004-10706}. It identifies ten tasks for data scientists to approach using the literature, which are shown in Table 1.


To test whether this set of tasks is has a consistent and complete explanatory power, we test whether it can explain the research goals identified by another source. The second source of COVID research goals that we use for this is the WHO's Roadmap to COVID-19 research \cite{WHO}, which proposes nine main research goals. We find that the Kaggle goals are a superset of the WHO goals. The WHO roadmap contains less goals, and these are more specific (e.g. the Kaggle `the vaccines and therapeutics' goal corresponds to the WHO's 'vaccines' and 'therapeutics' goals).

The data representation of a tasks depends on the (re)ranking method used to improve search results. These are based on a title and text description of approximately 220 words.

\subsection{Query-task mapping}
A manual and automatic task classification are compared where. In both approaches each query is classified into one task. The manual classification was performed by the first author by matching words in the topic fields to those in the task descriptions (with some liberty taken with regard to hypernyms and synonyms). In general, manually annotating search tasks based on a query alone can be difficult due to the ambiguous nature of search intentions, and thus may not be accurate. In the TREC setting however, the annotator is able to use the information available in the topic questions and narrative in addition to the query terms.

The results of the manual query-task mapping are shown in Table 1. We find that two tasks are represented by a significantly higher number of topics. These are tasks that contain actionable information for the searcher as an individual - how to prevent transmission, and how/when it could be treated. We compare the original 30 topics to the eventual set of 50 topics to see if the information needs changed over time. We notice that some tasks (e.g. transmission) were an immediate priority, whereas interest in other tasks (e.g. ethics) only later increased. One notable increase is in the genetics task. This reflects the comment online made by an organizer that, at some point, they realised that there were "not enough low-level biology topics" \cite{disco}.

\begin{table}
\begin{center}
\caption{Numbers of topics per task as identified using manual query-task mapping. Comparing the original 30 topics to the set of 50 topics months later.}
\begin{tabular}{ l | c c | l }
Task & 30T & 50T & Topic ID's\\
\hline
Trans, incubation, env & 7 & 10 & 1, 3, 12, 13, 14, 15, 20, 43, 46, 48 \\
Risk factors & 5 & 6 & 19, 21, 22, 23, 24, 40\\
Genetics, origin, evolution & 1 & 6 & 0, 30, 31, 35, 36, 39\\
Vaccines, therapeutics & 5 & 12 & 2, 4, 27, 28, 29, 32, 33, 37,\\
&&&38, 41, 45, 49\\
Medical care & 2 & 2 & 10, 16\\
Non-pharm. interventions & 4 & 5 & 9, 11, 17, 18, 47\\
Diagnostics, surveillance & 1 & 1 & 8\\
Geographic spread & 5 & 5 & 5, 6, 7, 25, 26\\
Ethical social consider. & 0 & 2 & 42, 44\\
Information sharing & 0 & 1 & 34\\
\end{tabular}
\end{center}
\end{table}

V{ö}lske et al. compared methods for query-task mapping based on query logs. They found that the most effective method was to index tasks in a small search engine, and then rank these tasks by a query using BM25. A query is classified as belonging to the top ranking task. We compared this approach to manual annotation and find a 66\% agreement between the methods, consistent with the findings of V{ö}lske et al. Because the agreement is low and we wish to focus on the potential of our task-based approach, we will focus on the manual annotation for the remainder of the paper.



\subsection{Improving search rankings}
We first introduce the baseline method, and then introduce three approaches to task-based ranking. The most promising of these approaches was query expansion based on task terms.


\subsubsection{Anserini baselines}
The baselines we use to test our research questions during the main experiment are based on those prepared by that the Anserini team each round \cite{DBLP:journals/corr/abs-2007-07846}\footnote{Further details available at \url{https://github.com/castorini/Anserini/blob/master/docs/experiments-COVID.md}, accessed August 2020}. We consider two variants of the Anserini baseline:
\begin{itemize}
    \item The \textbf{query} run ranks documents by BM25 on three different indices (full-text, title+abstract, title+abstract+paragraph). The scores for a document in those three rankings are combined using reciprocal ranking. The query consists of only topic query terms \footnote{This differs slightly from the fusion1 runs submitted by Anserini, as that was a concatenation of the query + question terms}.
    \item The \textbf{query+udel} run does the same, but uses a query generated by the University of Delaware's (udel) query generator. The query generator appends a topic's query and question attributes and then filters all terms which are not biomedical entities. These are identified using ScispaCy \cite{DBLP:journals/corr/abs-2007-07846}. This approach was a success in the early TREC-COVID rounds.
\end{itemize}

One interesting effect of combining these topic fields is that some terms are duplicated, which weights them more during the rankings. The query+udel run performed well in the TREC-COVID track since the early rounds (submitted as the fusion2 run).

\subsubsection{Re-ranking using Doc2Vec}
The \textbf{doc2vec} approach to model task context involves training a doc2vec model on both the paper abstracts and task descriptions (which are of similar length). Results are then re-ranked based on a linear combination of their BM25 score and their proximity to the task description vectors. This was tuned on the relevance assessments of the first round, as this was an early submission.

\subsubsection{Re-ranking by journals}
The journal-based approach re-ranks papers based on the journals they appeared in. Two variants were explored. In the \textbf{journal.prior} version, a prior likelihood is computed that papers from a given journal are relevant. This is based on the proportion of papers from a given journal were relevant in previous rounds. The likelihood is then normalized such that journals with only irrelevant papers get a score of -1, and journals with only relevant papers get a score of 1. Journals without prior information get a score of 0.

The task-dependent \textbf{journal.task} variant is similar. The same procedure is repeated for each task, but this time only using relevance assessments of topics that were manually classified into the current task. The task-based prior scores have some intuitive validity - some high scoring journals for the `risk factors' task include journals about diabetes and cardiovascular research. These are indeed some of the risk factors in the task description.

In both variants we calculate the relevancy score for each search result as a linear combination of the bm25 ranking and the journal's score. Tuning was performed on the cumulative relevance assessments of the first two rounds, as this run was an early submission to the TREC-COVID track.

\subsubsection{Task term expansion}
The task-based approach is to perform query expansion using task terms.  There is a \textbf{query+task} variant and a \textbf{query+udel+task} variant. The difference is that the latter includes topic question terms in addition to topic query terms.

Task terms were selected from the task description. First, biomedical entities are extracted from the task descriptions using ScispaCy's biomedical entity recognition. In order to keep the query short and specific, a selection of these task terms is made based on their TF-IDF score. This is based on the TF in the task description and the IDF in the collection of paper abstracts. The top n terms are then used, and appended to the query string as new terms. In order to weight the query terms more than task terms, we add duplicates of the original query terms to the new query string. Choosing how many terms should be added, and how these should be weighted is done using the topics and cumulative relevance judgements of the first four rounds.


In order to compare results to the Anserini baseline runs we create similar fusion runs. This entails performing the same ranking with three different indices (full-text, title+abstract and title+abstract+paragraph), and cominging the three scores each search result gets. 

When we test this task term expansion using the three indices (using $n = 3$ task terms, and using three duplicates of each query term) we find that adding task terms works best with a full-text index ($NDCG@20 = 0.3668$) and a title+abstract index ($NDCG@20 = 0.3600$). This approach severly underperforms on a title+abstract+paragraph index ($NDCG@20 =  0.0169$). This may be because this approach is very sensitive to paragraphs that use a task term in a different context. Parameters were tuned using the full-text index, resulting in the scores in Table 2. The results indicate that we should use $n = 3$ task terms, and that these should be weighted less than query terms.

Because we find our task terms to perform best on the full-text and abstract indices, we create the fusion runs by using only the additional task terms when querying the full-text index.



\begin{table}
\begin{center}
\caption{Parameter selection based on NDCG@20 scores using only the Anserini full-text index and the full round 4 cumulative judgements. Columns indicate the number of task terms added, and rows how the terms are weighted relative to different term types.}
\begin{tabular}{ l | c c }
Term weighting & 3 terms & 5 terms\\
\hline
1 query : 1 question : 1 task & 0.2915 & 0.2915\\
2 query : 2 question : 1 task & 0.3452 & 0.2982\\
3 query : 3 question : 1 task & 0.3668 & 0.3362\\
\end{tabular}
\end{center}
\end{table}

\section{Experiments}
All experiments are performed with the cumulative relevance assessments of the TREC-COVID rounds 1 through 4.

\subsection{Doc2Vec re-ranking}
During the first round we generated a BM25 baseline (full-text index) and a re-ranked \textbf{doc2vec} run. The baseline ($NDCG@20 = 0.2490$) outperforms the re-ranked run ($NDCG@20 = 0.0964$). It seems that a proximity between task descriptions and paper abstract in doc2vec space does not imply a semantic similarity.

\subsection{Journal re-ranking}
During the third TREC-COVID round we used the Anserini r3.rf run as a baseline\footnote{The same query generator as the \textbf{query+udel} baseline we introduced, but uses the abstract index and relevance feedback of the previous rounds instead of fusing results from multiple indices.}, which was compared with the re-ranked \textbf{journal.prior} and \textbf{journal.task} runs. The baseline ($NDCG@20 = 0.5800$) slightly outperforms the approach based on journal priors ($NDCG@20 = 0.3228$) and easily outperforms the task-based alternative ($NDCG@20 = 0.5406$). The task-based variant of the journal performs much better than the variant based on journal priors. This suggests that there is no objectiveness good journal, but that it depends on the context of the information need. The tasks may have been able to capture some but not enough of this context.

\subsection{Term expansion}
During the main experiment we use the \textbf{query} and \textbf{query+udel} baselines, which let us investigate our research questions. Table 3 displays our results using task-based term expansion, and compare these to the term expansion based on topic question terms.

\begin{table}
\begin{center}
\caption{Comparing fusion runs with various query term selection on the cumulative round 4 assessments.}
\begin{tabular}{ l | c c c c }
                    &30 topics & & 45 topics\\
                    & NDCG@20 & MAP & NDCG@20 & MAP\\
                    \hline
query&              0.4290 & 0.1626 & 0.4316 & 0.1910\\
query+udel&         0.5073 & 0.2082 & 0.4956 & 0.2300\\
\hline
query+task&         0.4433 & 0.1620 & 0.4446 & 0.1893\\
query+udel+task&    0.4929 & 0.2080 & 0.4907 & 0.2293\\
\end{tabular}
\end{center}
\end{table}

We find that adding task terms to query terms marginally improves results on NDCG@20, and results in a slightly lower MAP (RQ1). When adding these to a query that already contains question terms the MAP decreases (RQ2). It appears results were not significantly improved. The tasks describe the search intent at a higher/more abstract level than the topic question. When considering the findings in TREC's precision medicine track we find a potential explanation, as it was shown that using hypernyms during term expansion has a negative effect on search rankings \cite{DBLP:conf/sigir/FaesslerOH20}.  
The search tasks we identified may be too generic, and having more specific tasks may improve the efficacy of our approach. The scores of the \textbf{query+udel} run show a clear potential for improvements available by formulating better queries.

The scores remained consistent, and even slightly improved as new topics were introduced. This suggests that the tasks identified are stable and complete enough to deal with new topics (RQ3). 

\section{Conclusions}
A successful strategy during early rounds of the TREC-COVID track was to extend a user's query with other relevant terms. We explored task-based search for the scientific COVID-19 literature, which allowed us to generate task terms that the user might not have entered. Our approach to task-based ranking involved three main steps. We 1) identified search tasks based on research goals in the scientific community; 2) classified search queries into those search tasks; and 3) adjusted search rankings based on the task context of a given search query.

First, potential search tasks were identified and gathered in a task framework based on Kaggle's initial COVID challenge and WHO (COVID) research roadmap. Second, queries were manually mapped into tasks. Automatic task classification was explored, using an approach based on retrieving task descriptions from a small search engine, but it was found to be unsatisfactory. Third and finally, three methods were compared to improve search rankings based on the task context. Of these, the most promising approach was a query expansion approach that added task terms selected from task descriptions. The terms added were the biomedical entity terms with the highest TF-IDF score.


We found that using too many task terms negatively affect the search ranking, possibly due to query drift. We also found that task terms should be weighted less than query terms, and that this approach does not work with a paragraph index.


Our approach slightly improved NDCG@20 scores compared to using only query terms (RQ1). Our approach to query generation did not yet reach the potential that others have shown when using terms from a topic's question field, although question terms may not be available in a real life search situation. Our approach is a step towards achieving similar scores without requiring users to input additional terms.

When combining our method with the query generator based on topic question terms we find that the combination is less than the sum of its parts. This suggests that the question-based terms and the task-based terms do not reflect independent aspects of the information needs (RQ2). This leads to the hypothesis that the tasks represent the same information needs as the topic question, but that the information needs at the task level are too generic. If we identify more specific search tasks we may achieve better results.

The scores of our approach did not drop as new topics were introduced, suggesting that the task categorization was stable and complete enough to deal with these topics (RQ3).





In conclusion, our approach to modelling search task context slightly improved results over the baseline. This may be because the search tasks we identified capture information needs at a level that is too generic to improve search results much. The TREC-COVID track demonstrated the value of formulating better queries, and we in turn demonstrated that search task context could play a role in this.





\section{Discussion}
The challenge re-affirmed the value of traditional, fundamental concepts of IR such as query and document representation, and learning from relevance feedback. One example is how well the SMART system performed \cite{DBLP:conf/naacl/BuckleySA93}, which employs older technologies that were tuned well, based on experience.



\subsection{Dead ends}
We explored a number of methods which did not make it into the final approach. In the interest of reporting negative results, we mention these here.

\subsubsection{Filtering non-COVID papers}
An alternative method we tried was to filter documents in the CORD-19 dataset that were not specifically about COVID-19, but instead about related topics such as older viruses. This is a significant portion of the dataset - a majority of the documents in the dataset were published pre-2019. Evidence Partners created a distilled dataset using a combination of user annotations and machine learning in order to remove duplicates and documents not about COVID-19\footnote{\url{https://www.evidencepartners.com/resources/COVID-19-resources/}}. Filtering our results using this approach made results slightly worse, which suggests that search results from related topics are important to the new COVID topics. 


\subsubsection{Task term selection}
Additional attempts to select task terms were based on 1) selecting terms from the documents that were relevant for a given task and 2) selecting words from the task titles rather than the full descriptions. Neither were as effective as the final method.

\subsubsection{Annotator classification confidence}

We developed two variant systems that incorporated the annotator's confidence score in a manual classification  (in three levels). The first variant only tailored rankings to task context when classification confidence was high, and in the second variant the annotator confidence was proportional to how strongly task context affected search rankings. Both variants lowered MAP and NDCG scores.

\subsubsection{Faceted task framework}
We briefly explored an alternative domain-specific task framework that identified a small set of orthogonal task facets, which would be used to represent a large number of specific tasks. This approach is inspired by the generic faceted task categorization put forward by Li and Belkin \cite{DBLP:journals/ipm/LiB08}. Two examples of this kind of facet are the level of infection (individual/population level) and the topic facet (virus, transmission, host). These facets could combine to form goals such as epidemiology tasks (population level, transmission) or physical infection (individual level, transmission). This line of research was abandoned as we did not find a successful method to (re)rank documents based on the facets of a task.

\bibliographystyle{unsrt}
\bibliography{biblio}

\end{document}